\documentclass[prl,twocolumn,showpacs]{revtex4}

\usepackage{graphicx}
\usepackage{amsmath}
\usepackage{amssymb}

\begin{document}
\def\a{\alpha}
\def\b{\beta}
\def\e{\varepsilon}
\def\d{\delta}
\def\l{\lambda}
\def\m{\mu}
\def\t{\tau}
\def\n{\nu}
\def\o{\omega}
\def\r{\rho}
\def\S{\Sigma}
\def\G{\Gamma}
\def\D{\Delta}
\def\O{\Omega}

\def\ra{\rightarrow}
\def\ua{\uparrow}
\def\da{\downarrow}
\def\pd{\partial}
\def\bk{{\bf k}}
\def\bp{{\bf p}}
\def\bn{{\bf n}}

\def\be{\begin{equation}}\def\ee{\end{equation}}
\def\bea{\begin{eqnarray}}\def\eea{\end{eqnarray}}
\def\nn{\nonumber}
\def\lb{\label}
\def\pref#1{(\ref{#1})}


\title{Boundary Friction on Molecular Lubricants: Rolling Mode?}

\author{V.M. Loktev$^1$ and Yu.G.~Pogorelov$^2$}
\affiliation{$^1$Bogolyubov Institute for Theoretical Physics, 
National Academy of Sciences of Ukraine, Metrologichna Str. 14-b, 
Kiev, 03143 Ukraine, $^2$CFP/Departamento de F\'{i}sica, Universidade 
do Porto, 4169-007 Porto, Portugal}

\date{\today }

\begin{abstract}
A theoretical model is proposed for low temperature friction between two
smooth rigid solid surfaces separated by lubricant molecules, admitting
their deformations and rotations. Appearance of different modes of energy
dissipation (by ''rocking'' or ''rolling'' of lubricants) at slow relative
displacement of the surfaces is shown to be accompanied by the stick-and-slip 
features and reveals a non-monotonic (mean) friction force {\it vs} external load
\end{abstract}


\maketitle

\section{\label{sec:1} Introduction}

In the modern tribology a still increasing interest is put to the studies of
wearless friction on atomically smooth surfaces \cite{bhu},\cite{mclel} as a
possibility to provide an information about the basic processes of energy
losses on microscopic level, important for the purposes of optimization in
many technological applications. This is also connected with the search for
best coating and lubricant materials. The principal physical picture,
usually considered in relation with boundary friction on few molecular
layers of lubricant liquid, is the sequence of ''freezing-melting''
processes on the lubricant, giving rise to discontinious (stick-and-slip)
displacement of sliding surfaces \cite{isr}. Recently a new theoretical
approach was proposed for microscopic sliding processes at extremely low
velocities of motion and upon a monolayer of lubricant atoms \cite{pog}, as
can be the case for friction force microscopy (FFM). Based on the adiabatic
formation of metastable states (similar to dislocations in usual deformed
crystals or defects in the Frenkel-Kontorova model \cite{fk},\cite{sok}) and
their following relaxation, this treatment shows how the (average) microscopic 
friction coefficient depends upon the material parameters of the contacting solids
and lubricants and also how stick-and-slip jumps with atomic periodicity can
develop in the microscopic friction force. It should be noted that a similar
microscopic mechanism of dislocation-assisted sliding was recently proposed
for contacting asperities in dry friction \cite{bhu2}.

Ultimately, with the discovery of almost spheroidal molecule of fullerene C$%
_{60}$ \cite{kro} (and/or cylindrical carbon nanotubes \cite{Ii}) a hope had
arosen to use such closed molecular structures, as ''free rotating''
lubricants, for considerable reduction of the friction coefficient. For
instance, a reduction of sliding friction coefficient was already discussed,
due to involvement of spinning motion of surfaces in contact \cite{far}. 
However, the FFM experiments with use of C$_{60}$ monolayers deposited
over atomically smooth solid surfaces brought some contradictory results 
\cite{bhu1}, \cite{lut},\cite{sch}. To get their better understanding, a
further theoretical insight is desirable on the elementary processes of
boundary friction.

The present communication is aimed on extention of the above mentioned
adiabatic approach to the processes of boundary friction which include the
internal degrees of freedom of the lubricant molecules. Within a simplest
model, we make an attempt to show that, due to the discrete atomic structure
of such a molecule, there are possible qualitatively different modes of slow
motion, either dissipative or non-dissipative, depending on the applied
external load on the contact. For dissipative modes, there are energy losses
resulting from stick-and-slip discontinuities, but these losses turn out
much lower then for similar processes at sliding solid surfaces upon atomic
lubricants. Besides, the mean value of friction force is found to be a
non-monotonic function of the external load.

\section{\label{sec:2} The Model}

Let us consider a two-dimensional model for the boundary friction system
which includes two semi-infinite atomic arrays, the ''solids'', with
identical triangular lattice structure and a spatial separation $d$ between
their srufaces, and clusters of four atoms, the ''molecular lubricants'',
confined between the solids (Fig. {\ref{Fig.1}). The distances between nearest
neighbour atoms in both solids and in clusters are supposed invariable 
\cite{note} (that means, corresponding to ''absolutely rigid'' bonds), and 
the bond length for the molecule equals to the lattice parameter $a$ for the 
solid surfaces.

\begin{figure}
\centering{
\includegraphics[width=6.cm, angle=0]{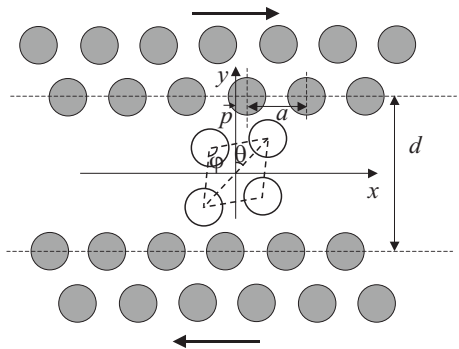}}
\caption{Schematic of the friction system: two solid arrays (grey circles)
with identical lattice structures are separated by a four-atomic lubricant
molecule (white circles). The molecule center rests at the origin, and
its rhombic configuration is determined by the angles $\theta$ and $\phi$,
while the symmetric displacements of solids are described by the parameter $p$
(see the text).}
\lb{Fig.1}
\end{figure}

The model includes certain ''soft links'', which determine the energy
variation {\it vs} relative displacement of the solids. In the first turn,
this is weak Lennard-Jones interaction between an atom of solid and a
lubricant molecule at distance $r$:

\begin{equation}
V_{LJ}(r)=\varepsilon _{0}\left[ \left( \frac{r_{0}}{r}\right) ^{6}-2\left( 
\frac{r_{0}}{r}\right) ^{12}\right] ,  \label{1}
\end{equation}
characterized by the adhesion energy $\varepsilon _{0}$ and equilibrium
distance $r_{0}$. Besides, a lubricant molecule can experience rhombic
deformation which changes its energy as $k\sin ^{2}\phi $, where $k$ and $%
\phi $ are the elastic constant and rhombic angle, respectively. We neglect
the interaction between different lubricants, that is omit their possible
collective modes. Thus the molecular lubricants are supposed to contribute
independently into the total friction force. Then we choose one particular
molecule and set its center of mass as the origin of reference frame, so its
configuration is determined by the ''internal'' rhombic angle, $\phi $, and
the ''external'' orientation angle, $\theta $ (say, with respect to the
normal to interface). In such a frame, the two solids are supposed to be
displaced symmetrically with respect to the lubricant molecule and their
configuration is fully defined by the displacement parameter $p$ chosen,
e.g., as the smallest positive longitudinal coordinate of the atoms of upper
solid.

All this permits to write the full energy (per one lubricant) in the simple
form: 
\begin{equation}
E=\sum_{n,m}V_{L-J}(|{\bf r}_{n}-{\bf R}_{m}|)+k\sin ^{2}\phi ,  \label{2}
\end{equation}
where ${\bf r}_{n}$ and ${\bf R}_{m}$ are respectively positions of atoms of
solids and atoms in the lubricant molecule. Obviously, this extremely
simplified model does not pretend to give a quantitative description and
explanation of friction experiments in real systems with complex and
typically incommensurate solid and molecular structures. It serves mainly to
illustrate some new qualitative possibilities for weakly dissipative
processes, associated with the internal (rolling or rotational) degrees of
freedom of molecular (nanotubes including) lubricants, in contrast to the
dissipation by only translational motions of the ''point-like'' (atomic)
lubricants.

\section{\label{sec:3} The Adiabatic Dynamics}

The adiabatic treatment of the system, corresponding to the expression (\ref
{2}), follows the lines suggested in Ref. \cite{pog}. The equilibrium
distance for Lennard-Jonnes interaction (\ref{1}) is taken equal to the
interatomic distance in the solids: $r_{0}=a$. At given separation $d$
between the surfaces, we calculate numerically the total energy profile $%
E(p,\theta ,\phi )$ as a function of the displacement parameter $p$, and
also of the angles $\theta $ and $\phi $. Next this function is optimized
with respect to the deformation angle $\phi $ to result in the profiles 
$E(p,\theta )$, such as displayed in Figs. \ref{Fig.2}-\ref{Fig.5} (for 
different values of $p$, through the whole displacement period from $0$ 
to $a$, and at different separations $d$).

The primary optimization in $\phi $ refers to the stronger elastic
deformation constant (we took $k=0.5\varepsilon _{0}$, while the amplitude
of relevant energy oscillations in $\theta $ is $\sim 0.1\varepsilon _{0}$),
and hence to a faster relaxation in $\phi $ than in $\theta $.\cite{note1} 
Then the system behavior at very slow uniform variation of the parameter $p$ 
with time (that is, the slow dynamics) is obtained from the analysis of the 
profile $E(p,\theta )$. Below we analyze how this profile changes with growing
external load, which is here simulated by a gradual decrease of the
separation distance $d$.\newline
{\it i}) At greatest separations (evidently corresponding to the lowest
loads), it is seen from Fig. \ref{Fig.2} that the energy profile has a single $\pi /2$
periodicity in $\theta $. This means that, for any given $p$, there is a
single equilibrium state for the molecule, characterized by its orientation $%
\theta \left( p\right) $ (within to a $C_{4}$ rotation) and energy $E\left(
\theta \left( p\right) \right) $ , such that $\left( \partial E/\partial
\theta \right) _{\theta \left( p\right) }=0$, $\left( \partial
^{2}E/\partial \theta ^{2}\right) _{\theta \left( p\right) }>0$. In this
case, the phase trajectory $E(\theta )$ (shown by the sequence of bold dots 
linked by arrows, for growing displacement) is closed and continious. Thus, the 
system energy changes in a fully reversible way and, though some forces are 
exerted in the process, their mean value over the cycle and so the mean friction 
force are exactly zero. This reversible variation of the angle $\theta $ around 
its median position (such that the long diagonal of the rhomb points vertically, 
inset to Fig. \ref{Fig.2}), corresponds to a ''rocking mode'' of the molecular motion.

\begin{figure}
\centering{
\includegraphics[width=6.cm, angle=0]{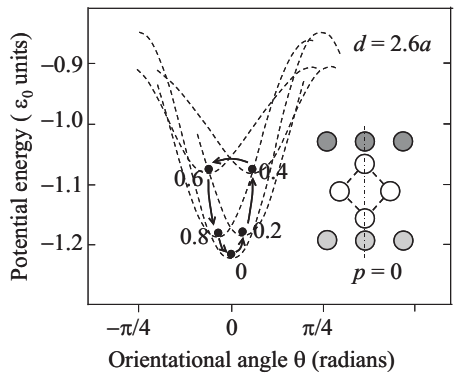}}
\caption{Series of profiles of potential energy {\it vs} orientation (at optimal
deformation, see the text) and the related equilibrium states (bold circles) of 
a lubricant molecule at growing displacement of the solids (the numbers 
indicate $p/a$ values) and at highest separation between them. The phase 
trajectory (arrows) shows that the molecule performs continuous "rocking" motions and 
after the full cycle (at $p/a=1$) returns to its initial state with no energy 
dissipation. Inset: the rhombic configuration at $p=0$.}
\lb{Fig.2}
\end{figure}

\noindent{\it ii}) When the two solid surfaces are getting closer (at increasing
load), the first qualitative change that appears in the system behavior is 
the doubling of its energy minima. Now there are two splitted minima at the
initial configuration $p=0$ (Fig. \ref{Fig.3}), which correspond to a two-fold
degenerate equilibrium state of the deformed lubricant molecule: the long
diagonal of the rhomb can deviate by a finite angle $<\pi /2$ to both sides
from the vertical (inset to Fig. \ref{Fig.3}). Since the molecule is considered a
classic object, it initially occupies only one of the minima (the left one
is chosen in Fig. \ref{Fig.3}).

\begin{figure}
\centering{
\includegraphics[width=6.cm, angle=0]{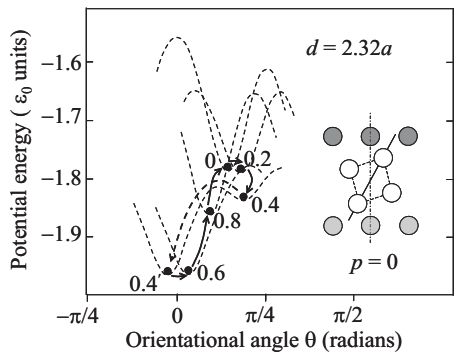}}
\caption{The same as in Fig. \ref{Fig.2} but at closer separation between the 
solids (higher external load). The lubricant molecule after "rocking" returns 
to its initial state, but in this course it experiences a thermally activated 
jump (the dotted arrow over the potential barrier for $p/a=0.4$) between the
metastable (0.4) and stable (0.4$^\prime$) minima, and the corresponding energy 
difference is irreversibly lost. Inset: the configuration at $p=0$ corresponds to 
one of the splitted minima}
\lb{Fig.3}
\end{figure}

However the degeneracy gets lifted for $p>0$, so that one of the splitted
minima turns stable ({\it s}) and another a metastable ({\it m}) equilibrium
state. As it is seen from the consecutive curves in Fig. \ref{Fig.3}, the energy 
barrier $h$ between the $m$-state and the nearest $s$-state (here that to the 
left from $m$) decreases by many times with growing $p$, as a certain function 
$h\left( p\right) $. Since the adiabatic lifetime for the $m$-state is: $\tau
_{m}=\tau _{a}\exp (\beta h)$ (where $\tau _{a}\sim 10^{-12}$ s is the
atomic oscillation time and $\beta $ is the inverse temperature), it
decreases in this course by many orders of magnitude. Eventually, this
lifetime gets comparable to the characteristic time $\tau _{0}$ of slow
displacement by an atomic period, at a very sharply defined instant when 
$p=p_{0}$ so that $h\left( p_{0}\right) =h_{0}=\beta ^{-1}\ln \left( \tau
_{0}/\tau _{a}\right) $. Hence it is almost exactly at this instant that a
thermally activated jump from $m$- to $s$-state is realized. After the jump,
the energy difference between $m$- and $s$-states (to the moment of
transition) is irreversibly lost, through the creation of quasiparticles
(phonons, for insulating solids) which are finally thermalized in the bulk.
For typical displacement rates in FFM $\sim 10^{2}$ ${\rm \AA }$/s, one has 
$\tau _{0}\sim 10^{-2}$ s, so that the barrier to the transition moment is
still as high as $h_{0}\sim 23\beta ^{-1}$.
\begin{figure}
\centering{
\includegraphics[width=6.cm, angle=0]{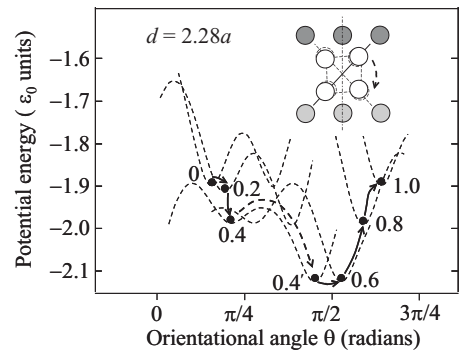}}
\caption{The same as in Figs. \ref{Fig.2} and \ref{Fig.3} but at still higher load. 
The thermally activated jump changes its direction and the lubricant molecule rotates 
by the angle $\pi/2$ after every period in $p$ ("rolling mode"). The energy loss 
and hence the friction force is somewhat larger than that in Fig. \ref{Fig.3}.}
\lb{Fig.4}
\end{figure}
For the situation presented in Fig. \ref{Fig.4}, the transition corresponds to 
$p_{0}=0.4a$ (the potential profile shown by the solid line). It is also seen
that, with further growing $p$, up to $a$, the system returns to its initial
state; thus the phase trajectory is still closed (''rocking mode'' again)
though discontinious. This discontiniuity produces stick-and-slip features
in the microscopic behavior of the force {\it vs} displacement and it is the
only source of the irreversible losses in our model system, in a full
similarity to the model of sliding solids upon atomic lubricants.\newline
There are however some distinctions between the two systems. Firstly, the
stick-and-slip profile for molecular lubricants is more complicated than the
simple triangular sawtooth for atomic lubricants. But especially important
is the fact that the irreversible forces for molecular lubricants are 
{\it smaller} than elastic, reversible forces (the jump heights in Figs. 
\ref{Fig.3}-\ref{Fig.5} are noticeably lower then the amplitudes of smooth 
oscillations), while the irreversible forces for atomic lubricants are orders 
of magnitude {\it higher} than the reversible ones. Since the reversible forces 
in both cases are characterized by the same energy scale $\varepsilon_{0}$, this 
indicates a possibility to essentualy reduce the dissipation by molecular (rotating)
lubricants {\it vs} that by atomic lubricants.
\begin{figure}
\centering{
\includegraphics[width=6.cm, angle=0]{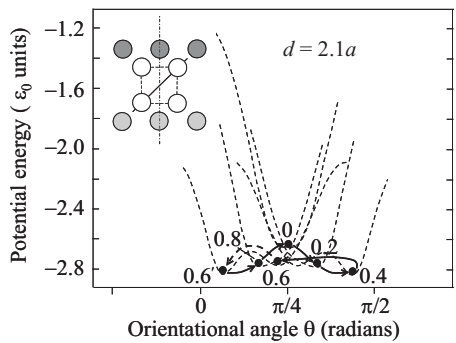}}
\caption{The same as in Figs. \ref{Fig.2}-\ref{Fig.4} at futher growing load. 
The molecule returns from "rolling" to "rocking" regime with an intermediate jump, 
but at much smaller energy loss than in Figs. \ref{Fig.3} and \ref{Fig.4}.}
\lb{Fig.5}
\end{figure}
Note, at least, that if the molecule occurs initially (at $p=0$) in the
right minimum, it simply spends a period, $0<p<a$, in the $s$-state, but
then passes to the $m$-state during the next period and eventually comes to the
same closed discontinuous rocking.\newline
{\it iii}) Now let us bring the surfaces yet a little bit closer, as by a
very small change of the distance $d$ between the solids (Figs. \ref{Fig.3} and 
\ref{Fig.4}). Then the system evolution acquires yet a new quality: now at the 
transition moment the nearest $s$-state is to the right from the given $m$-state. 
This results in that after the transition, at $p>p_{0}$, the molecule does not
return to its initial state, but will be eventually rotated by the angle 
$\pi /2$ (the dashed circles and arrow in the inset to Fig. \ref{Fig.4}), and 
this rotation will be repeated each next period. Thus, in this case we have
both unclosed (corresponding to the ''rolling mode'') and discontinious
regime of lubricant motion.\newline
{\it iv}) With still increasing loads, the situation of a single minimum for 
$E(0,\theta )$ will be restored again (though displaced by the angle $\pi /4$), 
corresponding to a ''square'' molecule (inset to Fig. \ref{Fig.5}). But now, unlike
the case of low loads, this minimum gets splitted with increasing
displacement $p$ (here at $p\approx 0.3$). The molecule returns from
''rolling'' to ''rocking'' regime (''rock-n-roll dance''), but with
considerably reduced energy dissipation: the energy loss at a jump for $d=2.1a$ 
(Fig. \ref{Fig.5}) is about 2.5 times smaller than for $d=2.28a$ (Fig. \ref{Fig.4}),
i.e. it {\it \ decreases} with load increasing. We do not present here detailed 
results of numerical simulations for even higher loads, resuming only that they 
reveal a number of subsequent dissipative regimes, either with growing and falling 
friction forces.
\section{\label{sec:4} Conclusions}
The above simple analysis demonstrates that boundary friction with
participation of molecular (spheroidal or cylindrical) lubricants can
possess quite unusual properties, such as existence of various regimes of
molecular motion, either non-dissipative and dissipative, with abrupt
transitions from one regime to another at continious variation of external
load. In the sequence of regimes (''rocking'' and ''rolling''), the obtained
friction force {\it vs} load displays non-monotonic and hence non-linear
behavior. This model (of course, with due improvements to be inserted) may
provide a mechanism for explaining the data of real experiments with the
fullerenes $C_{60}$ as lubricants and give indications for an optimal regime
of their practical applications.

\section*{Acknowledgments}

V.M.L. acknowledges partial support of this investigation by CRDF grant 
SCOPES 7UKPJ062150.00/1.


\begin{thebibliography}{99}
\bibitem{note} Here mean distances are meant, as the only relevant for adiabatic
equilibrium states, while small thermal fluctuations become important for
activation of relatively fast transitions between these equilibrium states.
\bibitem{note1}Similar hierarchy of relaxation rates between two degrees of 
freedom was also reported for point-like lubricants, corresponding to their 
normal (transversal) and longitudinal displacements \cite{pog}.
\end{thebibliography}
\end{document}